# Reverse Supply Chain Network Design of a Polyurethane Waste Upcycling System

Dalga Merve Özkan,[a] Sergio Lucia,[a] Sebastian Engell[a*]

[a]*Technische Universität Dortmund, Emil-Figge-Str. 70, 44227 Dortmund, Germany*
*sebastian.engell@tu-dortmund.de*

**Abstract**
This paper presents a general mathematical programming framework for the design and optimization of supply chain infrastructures for the upcycling of plastic waste. For this purpose, a multi-product, multi-echelon, multi-period mixed-integer linear programming (MILP) model has been formulated. The objective is to minimize the cost of the entire circular supply chain starting from the collection of post-consumer plastic waste to the production of virgin-equivalent high value polymers, satisfying a large number of constraints from collection quota to the quality of the feedstock. The framework aims to support the strategic planning of future circular supply chains by determining the optimal number, locations and sizes of various types of facilities as well as the amounts of materials to be transported between the nodes of the supply chain network over a specified period. The functionality of the framework has been tested with a case study for the upcycling of rigid polyurethane foam waste coming from construction sites in Germany. The economic potential and infrastructure requirements are evaluated, and it has been found that from a solely economic perspective, the current status of the value chain is not competitive with fossil-based feedstock or incineration. However, with the right economic incentives, there is a considerable potential to establish such value chains, once the upcycling technology is ready and the economic framework conditions have stabilized.

## 1. Introduction

In response to the urgent need for carbon-neutral economies and a more sustainable future, plastic waste upcycling has gained significant interest. Compared to mechanical recycling, this approach aims to extract valuable molecules from end-of-life or production waste via chemical processing, enabling their reintegration into the production of high-value polymers, and thereby fostering circular value chains. These value chains, with their potential to displace or diminish fossil-based raw material usage in polymer production, hold great promise for a sustainable future. However, design and implementation of such value chains is a demanding and complex task. For example, chemical upcycling often requires a specific type of waste feedstock, which must be carefully separated from other materials. Achieving these standards in waste separation can be logistically challenging and may necessitate substantial modifications to existing collection, sorting and dismantling processes. Secondly, the resource-intensive and complex nature of chemical upcycling technologies pose financial and operational hurdles, such as significant investments in technology and workforce training. Therefore, the successful integration of chemical upcycling technologies with waste management infrastructures and waste separation technologies requires the solution of a complex multi-faceted design problem to achieve regulatory compliance, and economic feasibility.

In this study, we present an adaptable framework that is designed to model, simulate, analyze, and optimize circular supply chains. We formulate the problem as a deterministic mixed-integer linear program (MILP), which computes strategic-level decisions on: (i) the optimal number, locations and sizes of the processing facilities, (ii) the optimal material flows between the nodes of the network under an economic objective. The multi-period deterministic model provides valuable insights into the layout of the system and the interactions among its components. It will be extended to a stochastic model to handle uncertainties in the future. Our framework can be applied to any region or supply chain of interest. We illustrate the proposed approach with the case study of a value chain for the upcycling of rigid polyurethane (PUR) foam waste in Germany as a representative of high-value plastic waste.

## 2. Model formulation

### 2.1. Sets

In the formulation, a node represents a geographical location in the studied region. We consider each node as a source of the targeted post-consumer waste material to be collected and as a possible location for installing facilities. The upcycling infrastructure model comprises six echelons: sources, collection facilities, recovery and treatment facilities, chemical processing facilities, downstream processing facilities, and consumers. We denote the set of sources by $S^o$. The set of materials, including intermediate and final products in the supply chain, is denoted by $P$. The sets of collection facilities, recovery and treatment facilities, chemical processing facilities and downstream processing facilities are denoted by $CF$, $RTF$, $CPF$ and $DPF$. The set of sinks (i.e. consumers of end products or chemical production facilities) of the upgrading system is denoted by $S^i$. For each type of facility there is a set of discrete size options denoted by $C^{CF}$, $C^{RTF}$, $C^{CPF}$ and $C^{DPF}$. The set of time periods is denoted by $T$.

### 2.2. Parameters

Each source $i \in S^o$ has a known waste supply $\sigma_{tpi} \in \mathbb{R}^+$ for a certain material type $p \in P$ in time period $t \in T$. Similarly, each consumer $n \in S^i$ has a demand $\delta_{tpn} \in \mathbb{R}^+$ for a certain product type $p \in P$ in time period $t \in T$. The minimum total processing quota (an environmental policy parameter) for a certain material type $p \in P$ that has to be collected from the sources in time period $t \in T$ is $\eta_{tp}$. The transportation cost associated with carrying one ton of material $p \in P$ per unit distance between the nodes is $t_p$. The transportation distances between the network nodes $i \in S^o$, $j \in CF$, $k \in RTF$, $l \in CPF$, $m \in DPF$ and $n \in S^i$ are represented by $D_{ij}$, $D_{jk}$, $D_{kl}$, $D_{lm}$ and $D_{mn}$, respectively. All transportation is assumed to be carried out via roads, and the transportation distances are estimated according to the *Haversine distance* formula. Each type of facility has a maximum capacity $\theta^c_{CF}$, $\theta^c_{RTF}$, $\theta^c_{CPF}$, $\theta^c_{DPF} \in \mathbb{R}^+$ for handling all materials $p \in P$ according to the choice of the size and an annualized installation cost $\alpha^I_{CF,c}$, $\alpha^I_{RTF,c}$, $\alpha^I_{CPF,c}$, $\alpha^I_{DPF,c} \in \mathbb{R}^+$ for each size option, an operating cost per ton of processed material $\alpha^O_{CF}$, $\alpha^O_{RTF}$, $\alpha^O_{CPF}$, $\alpha^O_{DPF} \in \mathbb{R}^+$ and a yield factor $\gamma_{p,CF}$, $\gamma_{p,RTF}$, $\gamma_{p,CPF}$, $\gamma_{p,DPF} \in \mathbb{R}$ for certain product $p \in P_{CF}^{out} \subset P, p \in P_{RTF}^{out} \subset P, p \in P_{CPF}^{out} \subset P, p \in P_{DPF}^{out} \subset P$ to be produced from a subset of materials $p \in P_{CF}^{in} \subset P, p \in P_{RTF}^{in} \subset P, p \in P_{CPF}^{in} \subset P, p \in P_{DPF}^{in} \subset P$.

## 2.3. Decision variables

In the presented model, there are two types of decision variables. The flows of a certain material type $p \in P$ between the nodes of the network are represented by a continuous variable $x \in \mathbb{R}^+$: The flow of material transported from source $i \in S^o$ to sink $j \in CF$ of capacity $c \in C^{CF}$ in time period $t \in T$ is $x_{tpijc}$, similarly from $j \in CF$ to $k \in RTF$ it is $x_{tpjkc}$, from $k \in RTF$ to $l \in CPF$ it is $x_{tpklc}$, from $l \in CPF$ to $m \in DPF$ it is $x_{tplmc}$, from $m \in DPF$ to $n \in S^i$ it is $x_{tpmn}$. The installation decisions of facilities are represented by binary variables $b \in \{0,1\}$ and can be stated as follows: The installation decision of $j \in CF$ of capacity $c \in C^{CF}$ is $b_{jc}$, similarly for $k \in RTF$ of capacity $c \in C^{RTF}$ it is $b_{kc}$, for $l \in CPF$ of capacity $c \in C^{CPF}$ it is $b_{lc}$, for $m \in DPF$ of capacity $c \in C^{DPF}$ it is $b_{mc}$.

## 2.4. Objective function

The objective is to minimize the total cost of the upcycling infrastructure. The first and second terms in Eq. (1) account for the installation and operating costs of facilities, and the last term accounts for the transportation costs of round trips.

$$
\begin{aligned}
min \Bigg\{ \Bigg( & \sum_{c \in C^{CF}} \sum_{j \in CF} \alpha_{CF,c}^{I} b_{jc} + \sum_{c \in C^{RTF}} \sum_{k \in RTF} \alpha_{RTF,c}^{I} b_{kc} + \sum_{c \in C^{CPF}} \sum_{l \in CPF} \alpha_{CPF,c}^{I} b_{lc} \\
& + \sum_{c \in C^{DPF}} \sum_{m \in DPF} \alpha_{DPF,c}^{I} b_{mc} \Bigg) + \sum_{t \in T} \Delta T_t \Bigg( \sum_{j \in CF} \alpha_{CF}^{O} \sum_{c \in C^{CF}} \sum_{p \in P} \sum_{i \in S^o} x_{tpijc} \\
& + \sum_{k \in RTF} \alpha_{RTF}^{O} \sum_{c \in C^{RTF}} \sum_{p \in P} \sum_{j \in CF} x_{tpjkc} + \sum_{l \in CPF} \alpha_{CPF}^{O} \sum_{c \in C^{CPF}} \sum_{p \in P} \sum_{k \in RTF} x_{tpklc} \\
& + \sum_{m \in DPF} \alpha_{DPF}^{O} \sum_{c \in C^{DPF}} \sum_{p \in P} \sum_{l \in CPF} x_{tplmc} \Bigg) + 2 \times \sum_{t \in T} \Delta T_t \Bigg( \sum_{c \in C^{CF}} \sum_{p \in P} \sum_{i \in S^o} \sum_{j \in CF} D_{ij} t_p x_{tpijc} \\
& + \sum_{c \in C^{RTF}} \sum_{p \in P} \sum_{j \in CF} \sum_{k \in RTF} D_{jk} t_p x_{tpjkc} + \sum_{c \in C^{CPF}} \sum_{p \in P} \sum_{k \in RTF} \sum_{l \in CPF} D_{kl} t_p x_{tpklc} \\
& + \sum_{c \in C^{DPF}} \sum_{p \in P} \sum_{l \in CPF} \sum_{m \in DPF} D_{lm} t_p x_{tplmc} + \sum_{p \in P} \sum_{m \in DPF} \sum_{n \in S^i} D_{mn} t_p x_{tpmn} \Bigg) \Bigg) \Bigg\}
\end{aligned}
\quad (1)
$$

## 2.5. Constraints

The following constraints are added to the problem. Demand satisfaction:

$$\sum_{m \in DPF} x_{tpmn} \leq \delta_{tpn} \quad \forall\, t \in T, p \in P, n \in S^i \quad (2)$$

The minimum collection quota is described as:

$$\sum_{c \in C^{CF}} \sum_{i \in S^o} \sum_{j \in CF} x_{tpijc} \geq \eta_{tp} \sum_{i \in S^o} \sigma_{tpi} \quad \forall\, t \in T,\, p \in P \quad (3)$$

The flow conservation at the sources can be defined as:

$$\sum_{c \in C^{CF}} \sum_{j \in CF} x_{tpijc} \leq \sigma_{tpi} \quad \forall \, t \in T, p \in P, i \in S^o \tag{4}$$

The flow conservation at the facilities (written for all other facilities similarly) is:

$$\gamma_{p,CF} \sum_{c \in C^{CF}} \sum_{p \in P_{CF}^{in}} \sum_{i \in S^o} x_{tpijc} = \sum_{c \in C^{RTF}} \sum_{k \in RTF} x_{tpjkc} \quad \forall \, t \in T, p \in P_{CF}^{out}, j \in CF \tag{5}$$

The maximum treatment capacity at the facilities (written for all other facilities similarly):

$$\sum_{p \in P} \sum_{i \in S^o} x_{tpijc} \leq \theta_{CF}^c \, b_{jc} \quad \forall \, t \in T, j \in CF, c \in C^{CF} \tag{6}$$

The selection of a single facility of same type (written for all other facilities similarly):

$$\sum_{c \in C^{CF}} b_{jc} \leq 1 \quad \forall \, j \in CF \tag{7}$$

The demand satisfaction constraint in Eq. (2) imposes the compliance with the capacities of the consumers. The constraints given in Eq. (3) and (4) ensure that the waste material is collected from the sources and shipped to collection facilities by respecting both the available waste material amount at the sources and the minimum collection quota. Eq. (5) enforces flow conservation at the facilities so that all the material entering a facility is processed and shipped to the next stage in the supply chain according to the yield factors associated with each technology and material. The maximum treatment capacities at the facilities given in Eq. (6) limit the total amount of material that can be delivered to a facility of chosen size. Eq. (7) makes sure that only one facility is chosen from multiple size (maximum capacity) options.

## 3. Case study description and assumptions

The proposed model is tested with a real case study for the upcycling of rigid polyurethane foam waste coming from construction sites in Germany. For more details about the estimation of annual PUR waste generation from construction materials and the selection of the nodes of the supply chain network, the reader can refer to Özkan et al. (2023). The potential locations for opening up facilities along with the relative amount of waste generated at the source nodes are shown in Figure 1a.

The supply chain operations are as follows: The PUR containing waste is collected separately in big bags on the construction sites, and not mixed with other construction waste. Then, all of the construction waste and the big bags are taken to CFs via skip trucks. The PUR waste is consolidated at CFs, and it is transported by trucks to RTFs for mechanical separation and compression into briquettes. After this stage, the briquettes are sent to CPFs where they are pre-conditioned and converted into pyrolysis oil via a catalytic pyrolysis process. Then the pyrolysis oil is further purified at DPFs to the desired final products. It is also possible to include the selection among technologies (or processing route) with the presented framework. However, the data on the PUR waste compositions as well as the details about the range of operations inside the facilities and their yields are not yet known precisely and still under research, therefore values that reflect the present state of knowledge are assumed and the selection of the processing routes selection is not considered in the current study.

In this study, for the CFs there are 8, for the RTFs, CPFs and DPFs there are 5 different processing capacity options each. The planning horizon includes three 4-month time periods, because the level of construction and demolishing or renovation activities in a region may vary. In this study, a variation in the volumes of the supplies by 10 to 30% is considered.

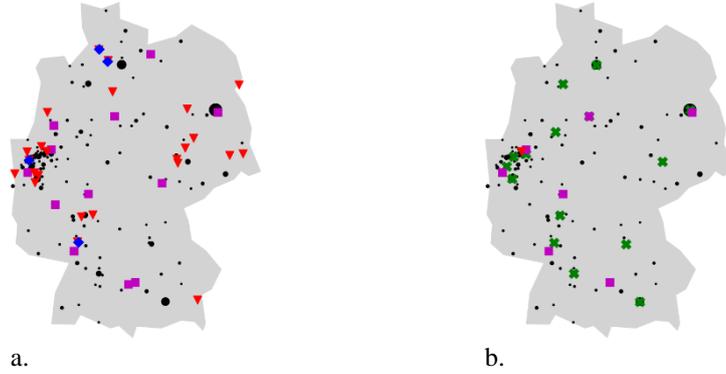

a.                    b.

**Figure 1.** a. The potential locations for opening up CFs (black circles), RTFs (magenta squares), CPFs and DPFs (red triangles) and the existing phosgenation plants to which the output of the chemical recycling can be fed (blue diamonds). The relative amount of waste material generated at the sources is indicated by the sizes of the circles. b. The optimal infrastructure layout with CFs (green crosses), RTFs (magenta squares), CPF and DPF (red triangle).

## 4. Results and discussion

The number of binary and continuous variables in the presented case study are 1284 and 2,789,388, respectively. The model is implemented using Python programming language and solved with Gurobi 10.0.3 to 3.5% optimality gap within 2.2 h on a computer with 13th Gen Intel® Core™ i9-13900K CPU @ 3.00 GHz and 128 GB RAM.

In the optimal layout, there are 15 CFs, 7 RTFs, one CPF and one DPF. The network has a decentralized structure at the collection stage; but becomes increasingly centralized over the following stages, culminating in a single, integrated chemical and downstream processing facility. The decentralized configuration of CFs and RTFs is primarily driven by the pressure to minimize the transportation costs associated with carrying the lightweight PUR material alongside other high-density construction waste (Each trip contains a very low wt. % of PUR waste). As the processing progresses along the value chain, the material undergoes transformations into denser forms – initially into briquettes and subsequently into pyrolysis oil. This densification significantly increases the transportation efficiency. Also, along the chain, the predominant cost factor shifts to the investment required for establishing technologically advanced chemical upcycling and downstream processing facilities. These facilities are capital-intensive, reaping the benefits of economies of scale. Consequently, establishing a single large-scale facility proves to be more economically viable than maintaining a decentralized structure. Supporting this conclusion, Ma et al. (2023) and Crîstiu et al. (2024) reported similar findings in their recent studies, which explored the reverse supply chain design for post-consumer plastic waste in the United States and Italy. The resulting optimal infrastructure design is shown in Figure 1b. The cost structures are shown in Figure 2 and it can be seen that the total cost is dominated by the investment and operating costs. In Figure 3, the material flows are visualized for a section of the supply chain with the help of Sankey diagrams.

## 5. Conclusions

This work introduces a general modelling framework that provides a basis for designing and understanding plastics upcycling value chains. The presented multi-period model enables the assessment of the system robustness in the presence of uncertain parameters, considering not only the changes in waste quantities but also changes in other parameter values, such as waste compositions, prices and regulatory constraints that may occur over time. Future work will include the extension to a stochastic model and the development of an independent actor model to better represent the multi-player nature of circular supply chains. Within this kind of model, money flows are also modelled and optimized, providing a basis for influencing the market by tuning the revenue streams and for developing market activating actions at the player-specific levels. This way, the right incentives that will drive the value chain towards a financially sustainable state can be determined and effective policies can be formulated.

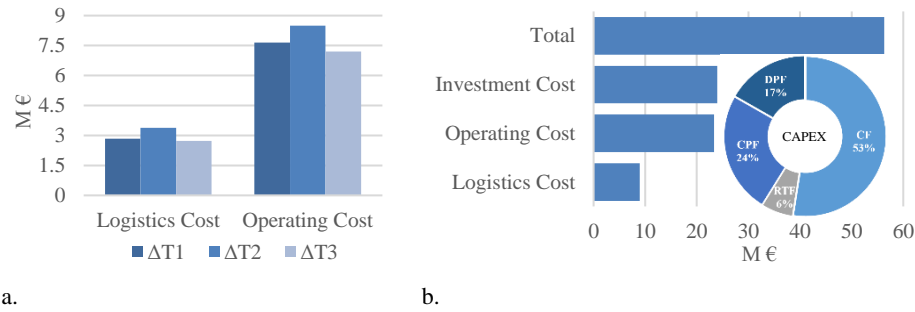

**Figure 2.** a. Logistics and operating costs in each time period, b. Total cost breakdown of the optimal solution.

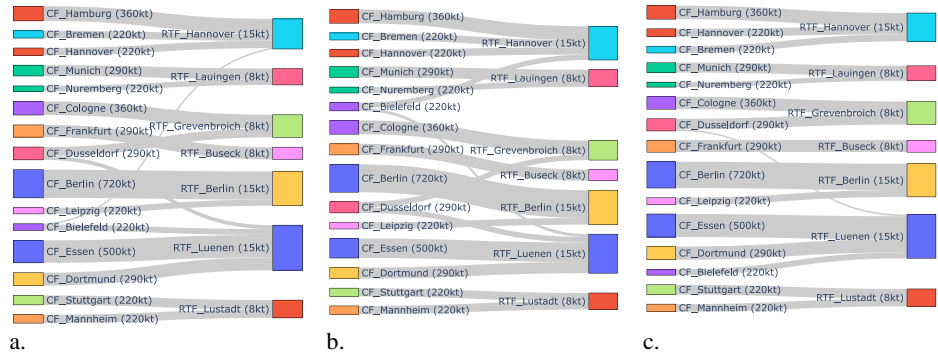

**Figure 3.** Sankey diagrams showing material flows from CFs to RTFs in a. $\Delta T_1$, b. in $\Delta T_2$, c. in $\Delta T_3$ (Annual maximum processing capacities of the facilities are shown in parentheses).

*Reverse Supply Chain Network Design of a Polyurethane Waste Upcycling System*

**Acknowledgments**

This study has funded by the European Union's Horizon 2020 research and innovation programme as part of the project "Circular Foam" under grant agreement No 101036854.